\documentclass[aps, prl, reprint, superscriptaddress, footinbib, twocolumn, showpacs]{revtex4-1}
\usepackage{amsmath} 
\usepackage{graphicx} 
\usepackage[OT4]{fontenc}

\usepackage{color}
\newcommand{\etal}[0]{{\em et al.}}

\newcommand{\Fig}[1]{Fig.~\ref{fig:#1}}

\begin{document}



\title{Cooling of a one-dimensional Bose gas}

\author{B.~Rauer}
\email[]{brauer@ati.ac.at} 
\affiliation{Vienna Center for Quantum Science and Technology, Atominstitut, TU Wien, Stadionallee 2, 1020 Vienna, Austria}

\author{P.~Gri\v{s}ins}
\altaffiliation[Present address: ]{INO-CNR BEC Center, Dipartimento di Fisica, Universit\`a di Trento, Via Sommarive 14, I-38123 Povo, Italy}  
\affiliation{Vienna Center for Quantum Science and Technology, Atominstitut, TU Wien, Stadionallee 2, 1020 Vienna, Austria}

\author{I.~E.~Mazets}
\affiliation{Vienna Center for Quantum Science and Technology, Atominstitut, TU Wien, Stadionallee 2, 1020 Vienna, Austria}
\affiliation{Wolfgang Pauli Institute, 1090 Vienna, Austria}

\author{T.~Schweigler}
\affiliation{Vienna Center for Quantum Science and Technology, Atominstitut, TU Wien, Stadionallee 2, 1020 Vienna, Austria}

\author{W.~Rohringer}
\affiliation{Vienna Center for Quantum Science and Technology, Atominstitut, TU Wien, Stadionallee 2, 1020 Vienna, Austria}

\author{R.~Geiger}
\altaffiliation[Present address: ]{SYRTE, Observatoire de Paris, 77 avenue Denfert Rochereau, 75014 Paris, France}  
\affiliation{Vienna Center for Quantum Science and Technology, Atominstitut, TU Wien, Stadionallee 2, 1020 Vienna, Austria}

\author{T.~Langen}
\email[]{tlangen@ati.ac.at}  
\altaffiliation[Present address: ]{JILA, NIST \& Department of Physics, University of Colorado, Boulder, Colorado 80309, USA}  
\affiliation{Vienna Center for Quantum Science and Technology, Atominstitut, TU Wien, Stadionallee 2, 1020 Vienna, Austria}
          
\author{J.~Schmiedmayer}
\affiliation{Vienna Center for Quantum Science and Technology, Atominstitut, TU Wien, Stadionallee 2, 1020 Vienna, Austria}

\date{\today}

\begin{abstract}
We experimentally study the dynamics of a degenerate one-dimensional Bose gas that is subject to a continuous outcoupling of atoms. Although standard evaporative cooling is rendered ineffective by the absence of thermalizing collisions in this system, we observe substantial cooling. This cooling proceeds through homogeneous particle dissipation and many-body dephasing, enabling the preparation of otherwise unexpectedly low temperatures. Our observations establish a scaling relation between temperature and particle number, and provide insights into equilibration in the quantum world.
\end{abstract}

\maketitle

\paragraph{Introduction.---\kern-1em}
Long coherence times, the tunability of many parameters and the ability to precisely probe and manipulate their quantum state make ultracold atomic gases a very  promising and versatile tool to study the physics of quantum many-body systems~\cite{Bloch2008}. The standard technique to reach the necessary ultracold temperatures in these systems is evaporative cooling~\cite{Hess1986,Masuhara1988,Luiten1996}, a fundamental process present in many physical systems from hot liquids to stellar clusters. It relies on the selective removal of the most energetic particles from a trapped gas and the subsequent re-thermalization of the gas to a lower temperature through elastic collisions. For efficient cooling this cycle is repeated continuously, increasing the phase-space density of a gas at the price of reducing the total number of atoms.

Intuitively, systems that do not thermalize can not be cooled through particle dissipation. In this letter we demonstrate that this notion is incomplete by studying the dissipative dynamics of a one-dimensional (1D) Bose gas with contact interactions. We observe a substantial decrease in temperature as a result of homogeneous particle dissipation, reaching temperatures far below $\hbar\omega_\perp$, the energy characterizing the transverse confinement. 
In this deep 1D regime thermalization is strongly suppressed~\cite{Kinoshita2006,Gring2012}: Thermalizing two-body collision are frozen out~\cite{Mazets2008} and three-body collisions~\cite{Mazets2008,Tan2010} or phonon-phonon scattering~\cite{Andreev1980,Stimming2011,Buchhold2015} can be neglected on our timescales. We find that the observed cooling can be modeled as a continuous density reduction extracting energy from the density quadrature of the free phononic excitations. Together with a continuous many-body dephasing this reduces the occupation number of each phonon mode and leads to a colder system. 

\paragraph{Experiment.---\kern-1em}
Our experimental system is a 1D Bose gas of $^{87}$Rb atoms prepared in the anisotropic magnetic trapping potential of an atom chip~\cite{Reichel2011}. Initially, a precooled three-dimensional (3D) cloud of thermal atoms, prepared in the $\rvert F,m_F \rangle \! = \,\rvert 2,2 \rangle$ ground state is loaded into the trapping potential. The gas is then cooled through the condensate transition and into the 1D regime by conventional evaporative cooling. The evaporation is realized through a weak driving of energy-selective radio-frequency (RF) transitions to untrapped Zeeman states~\cite{Davis1995}.  The cooling ramp of the RF frequency is performed over $1.6\,$s,  with the final part being particularly slow to minimize collective excitations in the gas. The 1D regime is reached when both the temperature and the chemical potential drop below $\hbar\omega_\perp$, the energy spacing between the ground state and the first excited state of the transverse confinement. At this point the condensate typically consists of $7000$ to $10000$ atoms at an initial temperature $T$ of $30$ to $100\,$nK. The resulting macroscopic quantum system is a quasi-condensate described by a macroscopic wave function with a fluctuating phase~\cite{Petrov2000}. 

\begin{figure}[!h]
  \centering
  \includegraphics[width=0.45\textwidth]{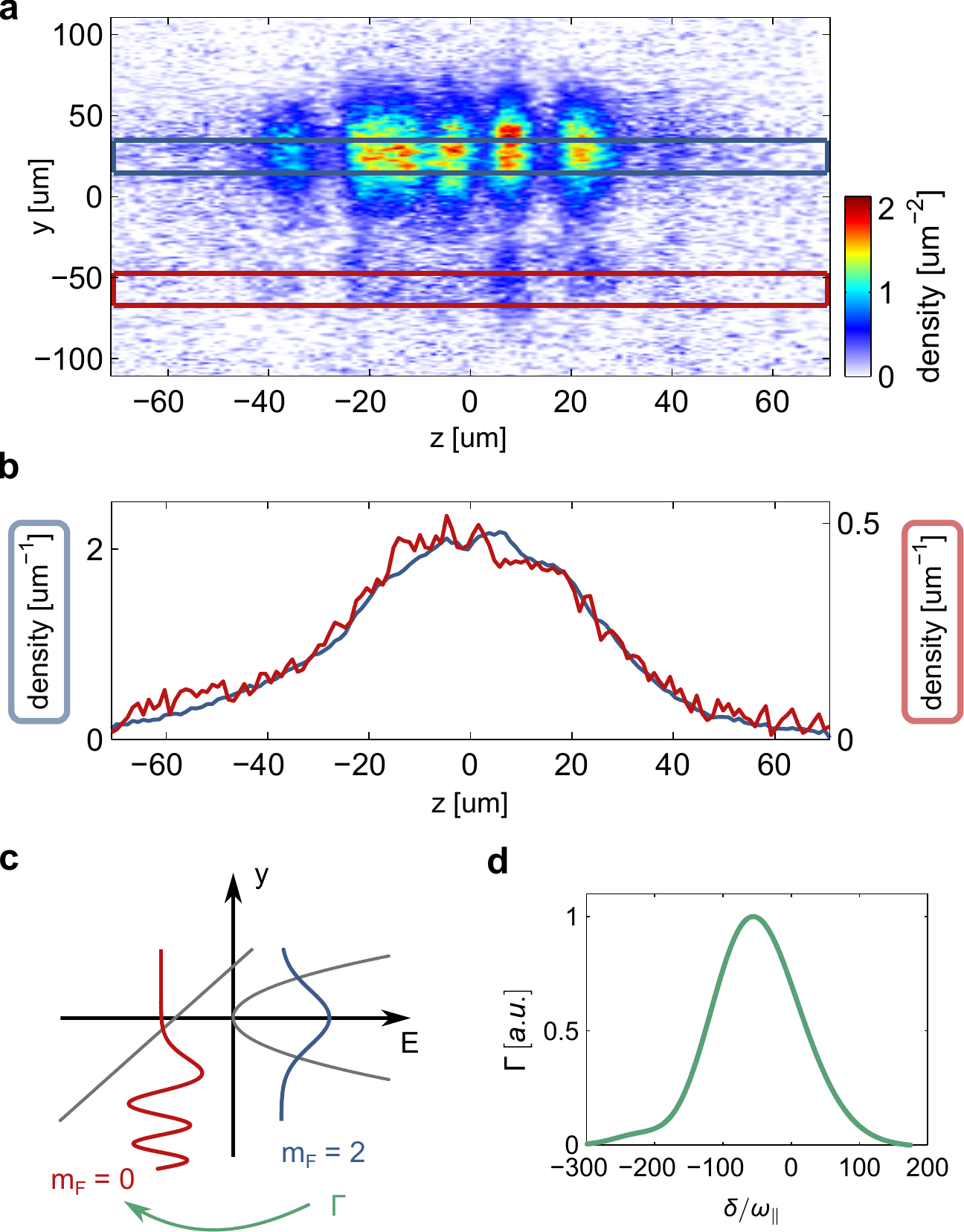}
  \caption{(color online). Analysis of the outcoupling mechanism. (a) A typical density speckle pattern formed by a 1D quasi-condensate expanding for $10.5\,$ms in time-of-flight. $2.5\,$ms before the gas was released a small fraction was outcoupled by a single $2\,$ms RF pulse. The outcoupled cloud is visible below the source cloud, showing the same speckle pattern. (b) Comparing the longitudinal density profiles of the source cloud (blue) and the outcoupled cloud (red) averaged over many realizations reveals the same form of the profile, indicating a homogeneous outcoupling process. (c) Sketch of the outcoupling mechanism. Transversal to the 1D axis the atoms occupy the ground state of the harmonic potential (blue). Outcoupling is achived by coupling these atoms to untrapped states, e.g. $\rvert 2,0 \rangle$ (red) which only feels the linear gravitational potential. The rate of this transition, taking only the transversal degrees of freedom into account, is plotted in (d) over the detuning of the driving field, given in units of the longitudinal trap frequency. The large width of this rate compared to the energy scale of the longitudinal dynamics leads to a nearly homogeneous outcoupling of atoms.
  }
  \label{fig:homogeneous_outcoupling}
\end{figure}

This trapped 1D quasi-condensate is the starting point of our experiments. To investigate the effects of particle dissipation we monitor the further evolution of the system under continuous driving of transitions to untrapped states. The outcoupling rate is tuned in such a way that atoms are removed slowly from the quantum degenerate gas to avoid collective excitations. In contrast to conventional evaporative cooling the atoms are not outcoupled energy-selectively but at a nearly homogeneous rate. This is experimentally demonstrated in \Fig{homogeneous_outcoupling}a by extracting a visible fraction of the gas using a single outcoupling pulse and analyzing its relation to the source cloud. The outcoupled atoms show the same average density profile as well as the same density speckle patterns, indicating a coherent and homogeneous outcoupling process.

Information about the system is extracted through absorption imaging in time-of-flight, transversal to the 1D axis~\cite{Smith2011}. To improve the individual images we employ a fringe removal algorithm~\cite{Ockeloen2010}. The imaging provides access to the evolution of the atom number, the density profile and the temperature of the gas. The latter is extracted from the longitudinal density speckle patterns forming in the time-of-flight expansion seen in \Fig{homogeneous_outcoupling}a~\cite{Imambekov2009,Manz2010}. These patterns are a direct result of the longitudinal phase fluctuations in the trapped gas. Extracting their normalized auto-correlation function and comparing it to simulated data generated through a stochastic Ornstein--Uhlenbeck process~\cite{Stimming2010} allows us to infer the temperature of the gas. 

\begin{figure}[tb]
  \centering
  \includegraphics[width=0.45\textwidth]{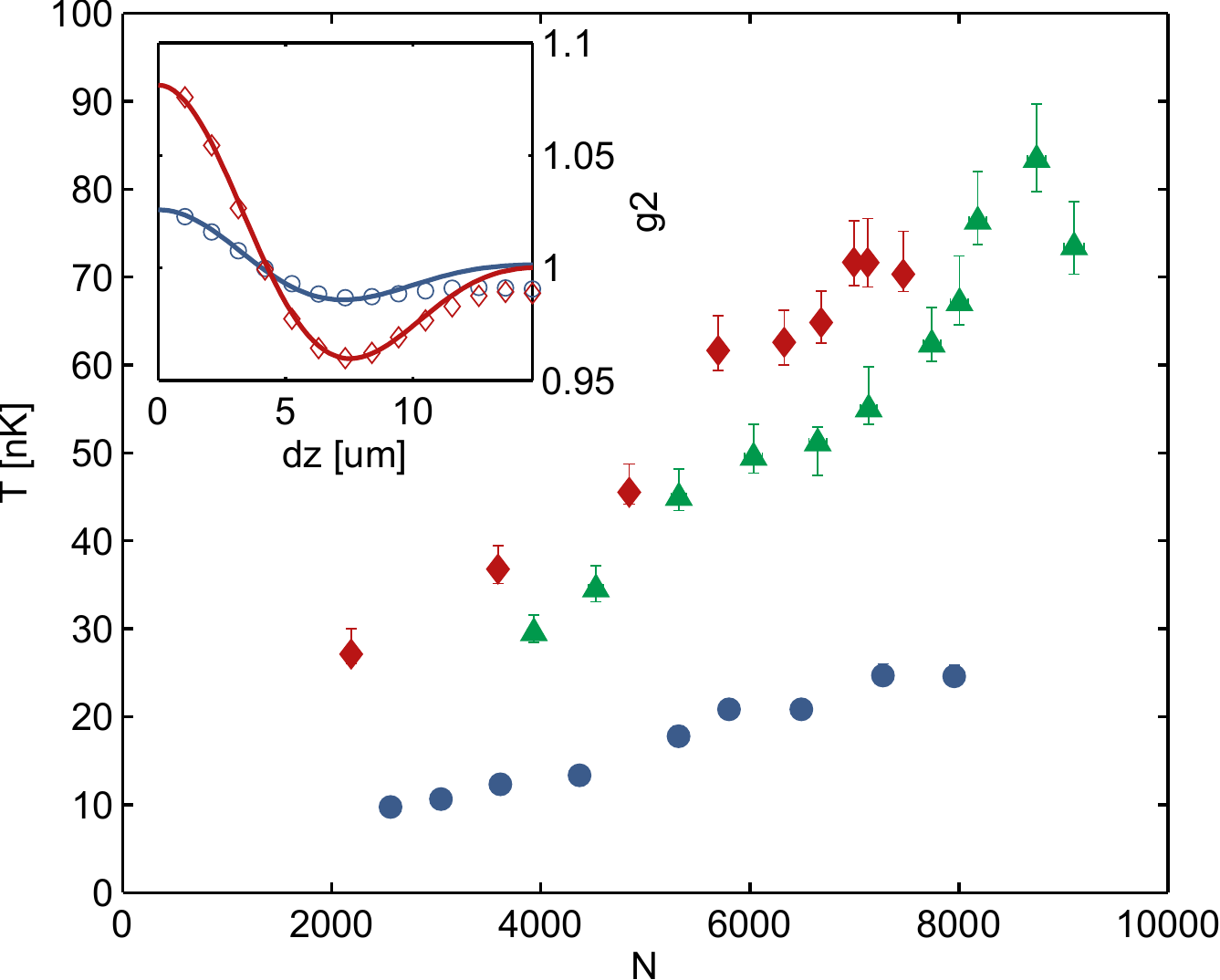}
  \caption{
  (color online). Three different examples for measurements of temperature $T$ over atom number $N$: The green triangles are obtained through a continuous RF-induced outcoupling from a harmonic trap with trap frequencies $\omega_\perp = 2\pi\times(2.1 \pm 0.1)\,$kHz and $\omega_\parallel = 2\pi\times(11 \pm 1)\,$Hz over a timescale of $25\,$ms. The blue circles differ in that they start from a much colder initial temperature and were measured over a timescale of $90\,$ms in a transversally slightly anharmonic trap created through a dressed state potential~\cite{Schumm2005}. There the trap frequencies are $\omega_\perp = 2\pi\times(1.6 \pm 0.1)\,$kHz and $\omega_\parallel = 2\pi\times(8 \pm 1)\,$Hz. Finally, the red diamonds were acquired through a very different measurement procedure, where MW radiation was used to couple atoms from the dressed trap directly to the anti-trapped $\rvert 1,1 \rangle$ state. Furthermore, in contrast to the other data points the temperature was measured after a constant period of outcoupling with a different outcoupling rate for each data point. The temperature error bars give the 68\% confidence interval obtained from a bootstrap. The inset depicts the speckle correlations of the coldest red and blue data points together with their respective thermal fits still showing good agreement.
}
  \label{fig:measurement}
\end{figure}

Our results for the evolution of the atom number and the temperature for different initial conditions, outcoupling rates and measurement procedures are shown in \Fig{measurement}. The outcoupling is achieved through RF transitions within the $F=2$ manifold or microwave (MW) transitions to the anti-trapped $\rvert 1,1 \rangle$ state. We find that in all cases the correlation functions of the density speckles in time-of-flight remain close to their thermal form for all data points. This is remarkable, as the system is close to an integrable point and thus can not thermalize on experimentally relevant timescales. In addition, the nearly homogeneous, energy independent out-coupling process would intuitively not lead to cooling. Nevertheless, we observe a significant decrease in temperature down to $k_BT \sim 0.1 \,\hbar \omega_\perp$. Moreover, even though the system constantly loses atoms the ratio $k_BT/\mu$ between the thermal energy 
and the interaction energy given by the chemical potential $\mu$ drops as well. In the coldest measurement presented in \Fig{measurement} (blue circles) it reaches values as low as $\sim \! 0.25$. The thermal coherence length $\lambda_T$ is inversely proportional to this ratio showing that the system becomes more coherent under dissipation. 

\begin{figure}[tb]
  \centering
  \includegraphics[width=0.45\textwidth]{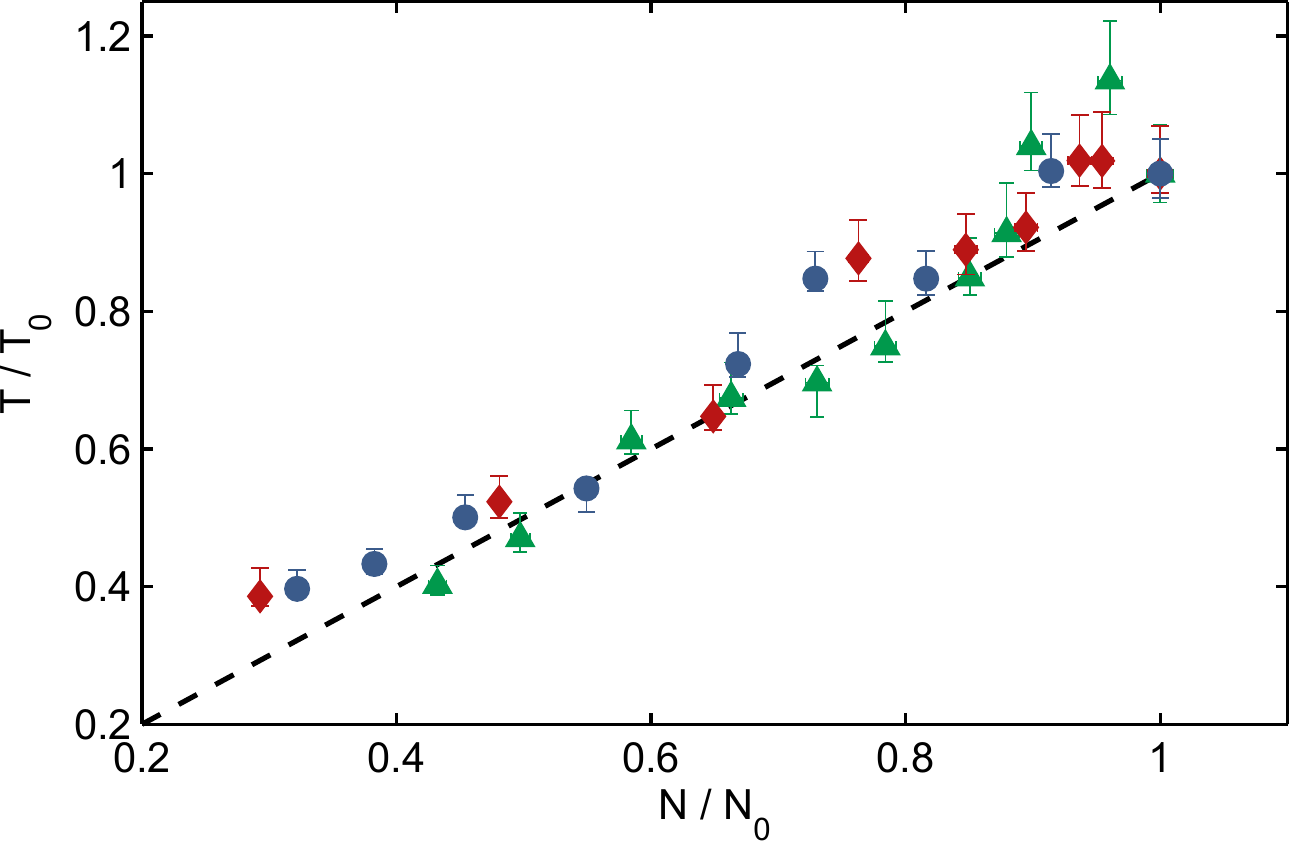}
  \caption{(color online). The data from \Fig{measurement} is plotted rescaled to their respective initial temperature $T_0$ and atom number $N_0$ values revealing a clear linear scaling.
}
  \label{fig:measurement_scaling}
\end{figure}

\Fig{measurement} suggests that the temperature decreases linearly with the number of atoms.  Rescaling the data points to their respective initial values $T_0$ and $N_0$ collapses the measurements to a single line, as shown in \Fig{measurement_scaling}. This suggests a linear scaling relation between temperature and atom number
\begin{equation}
	\frac{T}{T_0} = \frac{N}{N_0},
	\label{eq:scaling_eq}
\end{equation}
which is particularly interesting as these measurements are obtained from different experimental procedures and initial conditions.

\paragraph{Model.---\kern-1em}
In a pure 1D setting two-body collisions do not lead to a re-distribution of energy and momentum. However, in a quasi-1D trap with a tight transverse confinement this condition can be broken in collisions where there is enough energy available to access transverse excited states. These thermalizing two-body collisions are suppressed by a factor $\exp(-2 \hbar \omega_\perp/k_B T)$~\cite{Mazets2008} in a gas of non-degenerate bosons. For a degenerate gas the Bose enhancement of the ground state leads to an even stronger suppression. Consequently, these collisions freeze out as soon as the gas enters the 1D regime. Other processes that can lead to thermalization in our system are three-body collisions~\cite{Mazets2008,Tan2010} or phonon-phonon scattering~\cite{Andreev1980, Stimming2011, Buchhold2015}. However, their expected thermalization time-scales are beyond the times probed in our experiment and can not explain the observed cooling. A different case of evaporative cooling in a 1D harmonic trap has been modeled by Witkowska \etal \;\cite{Witkowska2011alternative}, assuming that atoms are outcoupled energy-selectively at the two ends of the cigar-shaped trap, which is not the physical situation in our experiment (see \Fig{homogeneous_outcoupling}).

To find a simple mechanism consistent with our experimental observations we start by considering the outcoupling process. The rate of outcoupling via RF or MW transitions is determined by the overlap of the condensate wavefunction with the free particle states of the untrapped Zeeman levels (see \Fig{homogeneous_outcoupling}c). The observed homogeneity can be explained by the fact that the energy of the longitudinal degrees of freedom is not resolved by this process since the timescale on which atoms leave the trap is determined by the much tighter transversal confinement (see \Fig{homogeneous_outcoupling}d). This means that each atom has the same probability to leave the trap irrespective of its position or energy. 

\begin{figure}[tb]
  \centering
  \includegraphics[width=0.45\textwidth]{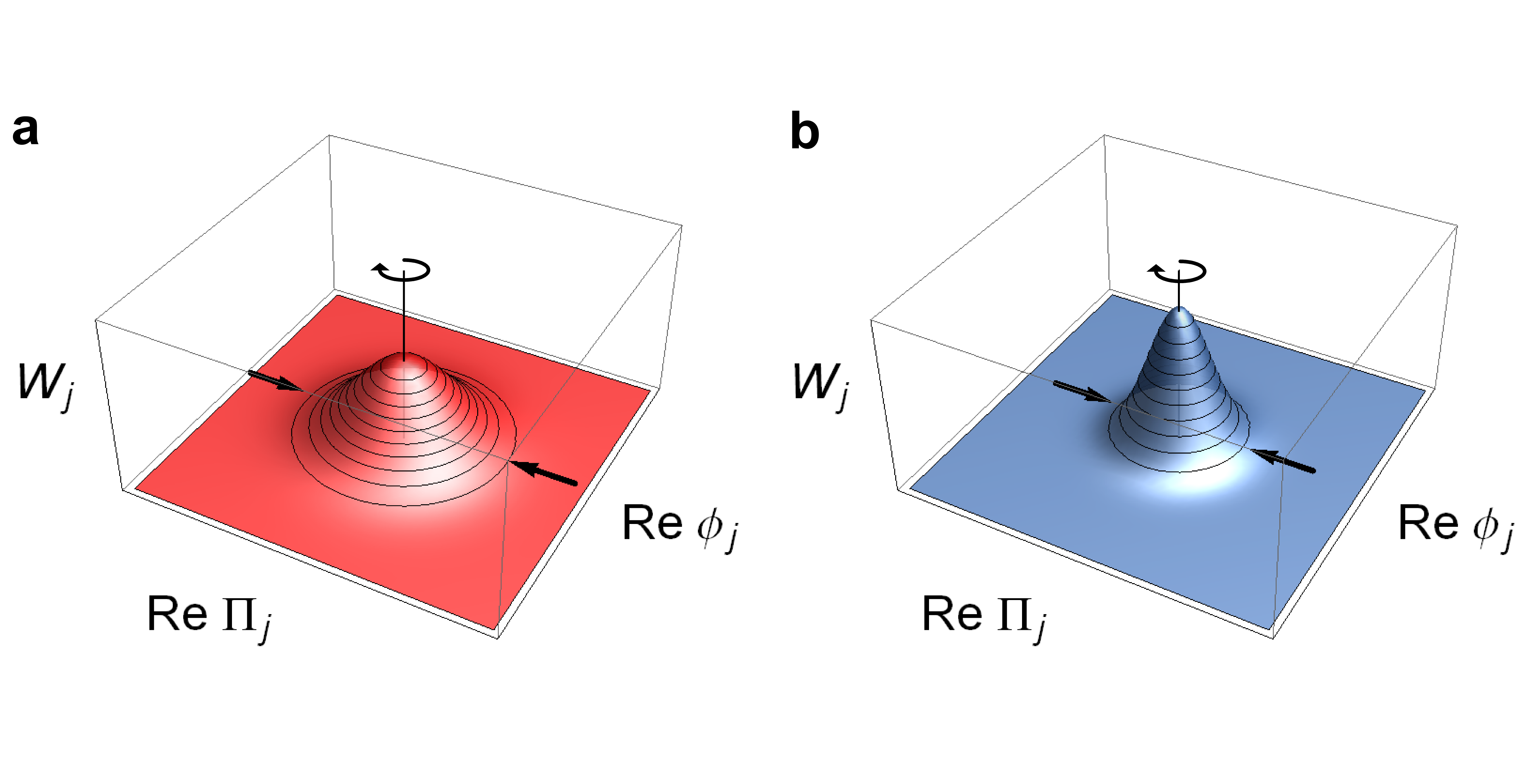}
	\caption{(color online). Dissipative cooling mechanism visualized through the Wigner function (a) of the phase and density quadrature $\phi_j$ and $\Pi_j$ of a single mode $j$, which rotates with the eigen frequency $\omega _j$. 
  The homogeneous outcoupling of atoms squeezes the Wigner function in the density quadrature, indicated by the black arrows, thereby reducing the energy of the mode. The rotation with $\omega _j$ together with a slow continuous squeezing leads to a (nearly) symmetric narrowing of the Wigner function (b) reflecting a reduced occupation of the mode. In our multimode system this happens in all modes at the same time, leading to a reduced temperature.
}
  \label{fig:wigner}
\end{figure}

The dynamics resulting from such a homogeneous particle dissipation can be intuitively understood within a Luttinger-liquid picture~\cite{Giamarchi2004}, describing the low energy dynamics of the underlying Lieb-Lininger model~\cite{Lieb1963}. The elementary excitations in this model are non-interacting phonon modes. Each of these modes contributes to the fluctuations in the gas through a density and a phase quadrature, in close analogy to the position and momentum quadratures of a harmonic oscillator. The free evolution of such a mode $j$ with energy $\hbar \omega_j$ can be visualized as a rotation of the corresponding Wigner function with the frequency $\omega_j$. In this picture, a sudden homogeneous outcoupling of atoms leads not only to a decrease in average density but also to a corresponding decrease in the density fluctuations. Such an instantaneous density reduction therefore extracts energy from the density quadrature of the phonon modes while leaving the phase quadrature unchanged. The system reacts to this reduction by dephasing, redistributing the remaining energy between the quadratures. A related process has recently been studied in detail using matter-wave interferometry after coherently splitting a 1D Bose gas and has been shown to lead to prethermalization~\cite{Gring2012,Kuhnert2013a} and light-cone-like spreading of thermal correlations~\cite{Langen2013}.

The experimentally investigated situation of a slow, continuous dissipation can be modeled as a series of small density reductions each followed by an immediate dephasing. This leads to a continuous decrease of the quasi-particle mode occupation, as illustrated in \Fig{wigner}. The mode occupations reacts to a sudden decrease in density $\rho \rightarrow \rho'$ as $n_j' = \frac{1}{2} \sqrt{\alpha}(\alpha^2 +1) \, n_j$, with $\alpha = \rho'/\rho$ being the ratio by which the average density is reduced. For a series of small reductions under continuous dephasing we can expand this factor around $\alpha = 1$, which, to leading order, gives $n_j' = \alpha^{3/2} \, n_j$. In the case of a harmonically trapped gas in the Thomas-Fermi limit where the central density scales as $\rho'/\rho = (N'/N)^{2/3}$ this leads to a linear scaling of $n_j$ with the particle number $N$. As the degenerate system is still sufficiently hot, the observable low-energy modes are dominated by the Rayleigh-Jeans part of the Bose-Einstein distribution where the mode occupation is proportional to the temperature $n_j \simeq k_B T / \epsilon_j$, with $\epsilon_j$ being the mode energy. Therefore, a homogeneous dissipation ultimately leads to a linear scaling relation between temperature and atom number $T/T_0 = N/N_0$, as observed in \Fig{measurement_scaling}. An extensive theoretical analysis of this process can be found in~\cite{Grisins2014}. Inhomogeneities in the outcoupling rate, stemming for example from the energy of the longitudinal degrees of freedom would lead to stable non-thermal energy distributions described by generalized Gibbs ensembles~\cite{Rigol2007,Langen2015}. 

This result extends scaling laws for the dynamical compression and expansion of a trapped 1D Bose gas~\cite{Rohringer2015} to the realm of open quantum systems. Its universality is a result of the phononic nature of the excitations. Supporting our simple model, the scaling law of Eq.\,\eqref{eq:scaling_eq} can also be found by numerically solving the Gross-Pitaevskii equation with an additional dissipative term. Furtheremore, it is interesting to note that this cooling mechanism is similar to cooling by adiabatic expansion, in that the entropy per particle stays constant. 

It has been suggested that vacuum fluctuations enter the system through the dissipation process~\cite{Grisins2014,Japha1999,Busch2014}, which would lead to a reduced cooling efficiency. However, all our measurements are consistent with the scaling of Eq.\,\eqref{eq:scaling_eq} derived from our simple model.

\paragraph{Conclusion.---\kern-1em}
We have experimentally observed the cooling of a degenerate atomic gas in the 1D regime by contious outcoupling of particles. In this process we reach temperatures as low as $k_BT \sim 0.25 \,\mu$ and $k_BT \sim 0.1 \,\hbar \omega_\perp$ which is far below the region where thermalizing two body collisions freeze out, rendering standard evaporative cooling ineffective. The observed cooling allows us to go deep into the 1D regime and is therefore of direct practical importance for all experiments with 1D Bose gases. Our model also explains how previous experiments could observe such low temperatures~\cite{Hofferberth2008alternative,Jacqmin2011}.  

Finally, our simple model suggests that the mechanism behind the observed cooling is not limited to 1D and could be relevant in 2D and 3D settings at very low temperatures where atoms are homogeneously outcoupled from the quantum degenerate gas and the dephasing of excitations is faster then their thermalization.

We would like to thank D. Adu Smith, M. Kuhnert and M. Gring for contributions in the early stage of the experiment, and J. Walraven, A. Polkovnikov, E. Demler and S. Weinfurtner for helpful discussions.  This work was supported by the Austrian Science Fund (FWF) through the project P22590-N16 and the SFB FoQuS Project F4010 and by the EU through the projects SIQS and the ERC advance grant QuantumRelax. B.R., P.G. and T.S. thank the FWF Doctoral Programme CoQuS (\textit{W1210}). R.G. acknowledges support by the FWF through the Lise Meitner Programme M-1423. T.L. acknowledges support by the Alexander von Humboldt Foundation through a Feodor Lynen Research Fellowship.

\bibliography{evaporative_cooling_books,Papers-EvaporativeCooling}

\end{document}